\documentclass[10pt,twocolumn,aps,pra,amsmath,amssymb,showpacs]{revtex4-1}
\usepackage{bm}
\usepackage{mathrsfs}
\usepackage{graphicx}
\usepackage[usenames,dvipsnames,svgnames,table]{xcolor}
\usepackage[unicode=true,
            pdfusetitle, 
            bookmarks=true,
            bookmarksnumbered=false,
            bookmarksopen=false,
            breaklinks=true,
            pdfborder={0 0 0},
            backref=false,
            colorlinks=true]{hyperref}
\hypersetup{linkcolor=NavyBlue,urlcolor=NavyBlue,citecolor=NavyBlue}

\usepackage{amsthm}
\newtheorem{prop}{\protect\propositionname}
\providecommand{\propositionname}{Proposition}
\newtheorem{theorem}{Theorem}

\usepackage{tikz-cd}

\newcommand{\cH}{\mathcal{H}}
\newcommand{\cS}{\mathcal{S}}

\begin{document}

\title{Markovianity of the reference state, complete positivity of the reduced dynamics, and monotonicity of the relative entropy}
\author {Iman Sargolzahi}
\email{sargolzahi@neyshabur.ac.ir; sargolzahi@gmail.com}
\affiliation {Department of Physics, University of Neyshabur, Neyshabur, Iran}

\affiliation{Research Department of Astronomy and Cosmology, University of Neyshabur, Neyshabur, Iran}
\author{Sayyed Yahya Mirafzali}
\email{y.mirafzali@vru.ac.ir}
\affiliation{Department of Physics, Faculty of Science, Vali-e-Asr University of Rafsanjan, Rafsanjan, Iran}

\begin{abstract}
Consider the set  $\mathcal{S}=\lbrace\rho_{SE}\rbrace$   of possible initial states of  the system-environment, \textit{steered} from a tripartite \textit{reference state}  $\omega_{RSE}$. Buscemi [F. Buscemi, \href{http://dx.doi.org/10.1103/PhysRevLett.113.140502} {Phys. Rev. Lett. {\bf 113}, 140502 (2014)}]  showed that the reduced dynamics of the system, for each $\rho_{S}\in \mathrm{Tr}_{E}\mathcal{S}$, is always completely positive if and only if $\omega_{RSE}$ is a \textit{Markov state}.
There, during the proof, it has been assumed  that the dimensions of the system and the environment can vary through the evolution.
Here, we show that this assumption is necessary: we give an example for which, though $\omega_{RSE}$ is not a Markov state, the reduced dynamics of the system is completely positive, for any evolution of the system-environment during which the dimensions of the system and the environment remain unchanged. 
As our next result, we show that the result of  Muller-Hermes and  Reeb [A. Muller-Hermes and D. Reeb,  \href{https://link.springer.com/article/10.1007%2Fs00023-017-0550-9}
 {Ann. Henri Poincare \textbf{18},  1777 (2017)}], of  monotonicity of the quantum relative entropy under positive maps, cannot be generalized to the  Hermitian maps, even within their physical domains.

\end{abstract}


\maketitle

\section{Introduction}

Consider a closed finite dimensional quantum system which evolves as
\begin{equation}
\label{eq:1}
\begin{aligned}
\rho\rightarrow\rho^{\prime} =\mathrm{Ad}_{U}(\rho)\equiv U\rho U^{\dagger},
\end{aligned}
\end{equation}
where $\rho$ and $\rho^{\prime}$ are the initial and final states (density operators) of the system, respectively, and $U$ is a unitary operator. ($UU^{\dagger}= U^{\dagger}U=I$, where $I$ is the identity operator.)

In general, the system is not closed and interacts with its environment. We can consider the whole system-environment as a closed quantum system which evolves as Eq. (\ref{eq:1}). So the reduced state of the system after the evolution is given by
\begin{equation}
\label{eq:2}
\begin{aligned}
\rho_{S}^{\prime}=\mathrm{Tr}_{E} \circ \mathrm{Ad}_{U}(\rho_{SE})=\mathrm{Tr}_{E}\left( U \rho_{SE}U^{\dagger}\right), 
\end{aligned}
\end{equation} 
where $\rho_{SE}$ is the initial state of the combined system-environment quantum system and $U$ acts on the whole Hilbert space of the system-environment. 

There was a tendency to assume that the reduced dynamics of the system can always be written as a completely positive trace-preserving (CP) map; i.e. it can be written as 
\begin{equation}
\label{eq:3}
\begin{aligned}
\rho_{S}^{\prime}=\sum_{i}E_{i}\,\rho_{S}\,E_{i}^{\dagger},\ \ \ \sum_{i}E_{i}^{\dagger}E_{i}=I_{S},
\end{aligned}
\end{equation}
where $\rho_{S}=\mathrm{Tr}_{E}(\rho_{SE})$ is the initial state of the system. In addition, 
 $E_{i}$ are linear operators and $I_{S}$ is the identity operator on the Hilbert space of the system $\cH_{S}$ \cite{1}.

 But, in general, this is not the case. In fact, the CP-ness of the reduced dynamics has been proven only for some restricted sets of initial states of the system-environment \cite{2, 3, 4, 5, 6,7,8}.
 
A remarkable result in this context is that given in \cite{6}. Consider the set  $\mathcal{S}$  of possible initial states of  the system-environment, \textit{steered} from a tripartite  state  $\omega_{RSE}$. There, it has been shown that, for all $\rho_{SE}\in\mathcal{S} $, the reduced dynamics of the system is CP, for arbitrary $U$, if and only if $\omega_{RSE}$ is a so-called \textit{Markov state}.

The above result is important, not only because it includes all its previous results \cite{7}, but also because it is, in fact, the most general possible result \cite{9}, at least, within the framework of \cite{10}. In the next section, we will review this result.

During the proof of the above result in \cite{6}, it has been assumed that the dimensions of the Hilbert spaces of the system $\cH_{S}$ and the environment $\cH_{E}$ can vary during the system-environment  evolution $U$, in general. In \cite{11}, we have questioned whether this assumption can be relaxed or not.  In Sec.  ~\ref{sec: C}, we show that this assumption is necessary for the result of  \cite{6}: we give an example, for which, though $\omega_{RSE}$ is not a Markov state, the reduced dynamics is CP, for any evolution $U$, which does not change $d_S$ and $d_E$, the dimensions of  $\cH_{S}$ and $\cH_{E}$, respectively.

We give our next result, on   monotonicity of  \textit{quantum relative entropy}, in Sec.  ~\ref{sec: D}.
The quantum relative entropy of the state $\rho$ to another state $\sigma$ is defined as 
\begin{equation}
\label{eq:4}
\begin{aligned}
S(\rho\vert\vert\sigma)= \mathrm{Tr}(\rho\mathrm{log}\rho)-\mathrm{Tr}(\rho\mathrm{log}\sigma),
\end{aligned}
\end{equation}
if $\mathrm{supp}[\rho]\subseteq \mathrm{supp}[\sigma]$, otherwise it is defined to be $+\infty$
\cite{1}.  ($\mathrm{supp}[\eta]$, the support of  the state $\eta$, is the subspace spanned by the eigenvectors of 
$\eta$ with nonzero eigenvalues.)

It was known that the relative entropy is monotone under CP maps $\Psi$ \cite{1, 12}:
\begin{equation}
\label{eq:5}
\begin{aligned}
S(\rho\vert\vert\sigma)\geq S(\Psi(\rho)\vert\vert\Psi(\sigma)),
\end{aligned}
\end{equation}
for arbitrary states $\rho$ and $\sigma$.
Recently, the above result has been generalized to the case of positive trace-preserving  maps, too \cite{13}; i.e., in Eq. (\ref{eq:5}),  $\Psi$ can be a  positive trace-preserving map. Positive maps are those which map any positive operator to a positive operator.
If we consider  positive maps as the most general physical time evolutions, then this result means that the relative entropy is monotone under any physical evolution. 

But, in \citep{14}, it has been shown that any Hermitian trace-preserving  map $\Phi$ is physical within its \textit{physical domain}. By a Hermitian map, we mean a map which maps any Hermitian operator to a Hermitian operator. Therefore,  $\Phi$ may  maps a state to a non-positive operator. So, in \citep{14}, the physical domain of  $\Phi$  is defined as the set of states which are mapped by $\Phi$  to density operators.

In Sec.  ~\ref{sec: D}, using a theorem proven in \cite{6}, we show that one can find physical evolutions, given by  Hermitian trace-preserving  maps $\Phi$, for which the relative entropy increases after the evolution. So, the result of \cite{13}  cannot be generalized to the Hermitian trace-preserving evolution, in general. In addition, we illustrate this result, using the example given in Sec.  ~\ref{sec: C}.

In the example considered in Sec.  ~\ref{sec: D}, $d_S$ and $d_E$
 vary through the evolution. 
In Sec.  ~\ref{sec: E}, we give another example in which the Hermitian non-positive evolution does not change $d_S$ and  $d_E$, but the monotonicity of relative entropy is again violated. This shows that the variation of $d_S$ and $d_E$ is not necessary for the non-monotonicity of relative entropy, under Hermitian non-positive evolution. 

Finally, in Sec. ~\ref{sec: S}, we end our paper, with a summary of our results.

\section{Reduced dynamics of an open quantum system} \label{sec: B}
\subsection{Reduced dynamics for a steered set} 

Consider the tripartite state  $\omega_{RSE}$,  on the Hilbert space of the reference-system-environment $\cH_{R}\otimes\cH_{S}\otimes\cH_{E}$, where $\cH_{R}$ is an ancillary Hilbert space.
The set of \textit{steered states} from performing  measurements on the part $R$ of  $\omega_{RSE}$ is \cite{6}
\begin{equation}
\label{eq:6}
\begin{aligned}
\mathcal{S}= \left\lbrace\rho_{SE}= \frac{\mathrm{Tr}_{R}[(P_{R}\otimes I_{SE})\omega_{RSE}]}{\mathrm{Tr}[(P_{R}\otimes I_{SE})\omega_{RSE}]} ,  P_{R}>0 \right\rbrace ,  
\end{aligned}
\end{equation}
where $P_{R}$ is  arbitrary positive operator on $\cH_{R}$ such that $\mathrm{Tr}[(P_{R}\otimes I_{SE})\omega_{RSE}]>0$ and $I_{SE}$ is the identity operator on $\cH_{S}\otimes\cH_{E}$. Note that, up to a positive  factor, $P_{R}$ can be considered as an element of a POVM. 

In Ref. \cite{6}, it has been shown that the reduced dynamics of the system, for all $\rho_{S}\in \cS_S\equiv\mathrm{Tr}_E\cS$ and arbitrary $U$, is CP if and only if  $\omega_{RSE}$ is a\textit{ Markov state}; i.e., if it can be written as \cite{15}
\begin{equation}
\label{eq:7}
\begin{aligned}
\omega_{RSE}=\mathrm{id}_{R}\otimes \Lambda_{S}^{CP} (\omega_{RS}),  
\end{aligned}
\end{equation}
where $\omega_{RS}=\mathrm{Tr}_{E}(\omega_{RSE})$, $\mathrm{id}_{R}$ is the identity map on $\mathcal{L}(\cH_{R})$ and $\Lambda_{S}^{CP}\,:\mathcal{L}(\cH_{S})\rightarrow\mathcal{L}(\cH_{S}\otimes\cH_{E})$ is a CP map. ($\mathcal{L}(\cH)$ is the space of linear operators on $\cH$.)

   When $\omega_{RSE}$  is  a Markov state, then  there exists a decomposition of the Hilbert space of the system $S$ as $\cH_{S}=\bigoplus_{k}\cH_{s_{k}}=\bigoplus_{k}\cH_{s^{l}_{k}}\otimes\cH_{s^{r}_{k}}$  such that
\begin{equation}
\label{eq:8}
\omega_{RSE}=\bigoplus_{k}\lambda_{k}\:\omega_{Rs^{l}_{k}}\otimes\omega_{s^{r}_{k}E},
\end{equation}
where $\lbrace \lambda_{k}\rbrace$ is a probability distribution ($\lambda_{k}\geq 0$, $\sum_{k}\lambda_{k}=1$), $\omega_{Rs^{l}_{k}}$ is a state on $\cH_{R}\otimes\cH_{s^{l}_{k}}$, and $\omega_{s^{r}_{k}E}$ is a state on $\cH_{s^{r}_{k}}\otimes\cH_{E}$ \cite{15}.

Let us summarize the result of this subsection, for later reference \cite{6}:
\begin{theorem}
\label{thm 1}
Assume that the set of possible initial states of the system-environment is given by $\cS$, in Eq. \eqref{eq:6}, which is 
 the steered set from a  tripartite state  $\omega_{RSE}$. The reduced dynamics of the system, in Eq. \eqref{eq:2}, is CP, for arbitrary $U$ and any  $\rho_{SE}\in \cS$, if and only if $\omega_{RSE}$ is a Markov state, as Eq. \eqref{eq:8}.
\end{theorem}

The following point is also worth noting. In this paper,  when we say that the reduced dynamics is given by a CP map $\Psi$, we mean that there exists a CP extension of $\Psi$  to the whole $\mathcal{L}(\cH_S)$, as Eq. \eqref{eq:3}, such that the reduced dynamics of the system, for each $\rho_{S}\in \cS_S$, is given by this CP map. 

 \subsection{Reduced dynamics for an arbitrary set} \label{sec:B-2}
 
A general  framework for linear (Hermitian) reduced dynamics, when both the system and the environment are finite dimensional, has been introduced in \cite{10}.
In this paper, we will restrict ourselves to the case that there is a one to one correspondence between the system initial states $\rho_{S}$ and the system-environment initial states $\rho_{SE}$.
So,  in this subsection, we review the framework of \cite{10} for this case.

Consider the set $\mathcal{S}=\lbrace\rho_{SE}\rbrace$ of possible initial states of the system-environment. Since, both the system and the environment are finite dimensional, a finite number $m$ of the members of $\mathcal{S}$, where the integer $m$ is  $0< m\leq {(d_S)}^2{(d_E)}^2$, are linearly independent. Let us denote this linearly independent set as
  $\mathcal{S}^\prime  = \lbrace\rho_{SE}^{(1)}, \rho_{SE}^{(2)}
  , \ldots\ , \rho_{SE}^{(m)}\rbrace$. Therefore, any $\rho_{SE}\in \mathcal{S}$ can be written as $\rho_{SE}=\sum_{i=1}^m a_i \rho_{SE}^{(i)}$, where $a_i$ are real coefficients.

We restrict ourselves to the case that all $\rho_{S}^{(i)}=\mathrm{Tr}_E(\rho_{SE}^{(i)}) \in \mathcal{S}^\prime_S\equiv \mathrm{Tr}_E\mathcal{S}^\prime$, $i=1, \dots ,  m\leq {(d_S)}^2$, are also linearly independent.
Therefore, there is a one to one correspondence between the members of $\mathcal{S}$ and the members of  $\mathcal{S}_S=\mathrm{Tr}_E\mathcal{S}$.
It is worth noting that  when $\mathcal{S}$ is a steered set, as Eq. \eqref{eq:6}, from a Markov state $\omega_{RSE}$, as Eq. \eqref{eq:7}, then the above correspondence holds \cite{9}.

 The subspace $\mathcal{V}$ is defined as the subspace spanned by $\rho_{SE}^{(i)} \in \mathcal{S}^\prime$ \cite{10}. 
  Therefore, each  $X\in \mathcal{V}$ can be expanded as $X= \sum_{i=1}^m c_i \rho_{SE}^{(i)}$, with complex coefficients $c_i$. In addition, for each $x=\mathrm{Tr}_E(X) \in \mathcal{V}_S \equiv\mathrm{Tr}_E\mathcal{V}$, we have $x= \sum_{i=1}^m c_i \rho_{S}^{(i)}$.
  
Let us denote the set of density operators in $\mathcal{L}(\mathcal{H}_S\otimes \mathcal{H}_E)$ and $\mathcal{L}(\mathcal{H}_S)$ by $\mathcal{D}_{SE}$ and $\mathcal{D}_{S}$, respectively.    
   Note that $\mathcal{S}\subseteq\mathcal{V}\cap\mathcal{D}_{SE} $ and $\mathcal{S}_S\subseteq\mathcal{V}_S\cap\mathcal{D}_{S}$.
 So,  that which we show for the whole $\mathcal{V}$ and $\mathcal{V}_S$ is also valid for their subsets $\mathcal{S}$ and $\mathcal{S}_S$, respectively.
 
Since all  $\rho_{S}^{(i)}\in\mathcal{S}_S^\prime$ are linearly independent, as all $\rho_{SE}^{(i)}\in\mathcal{S}^\prime$, for each $x\in \mathcal{V}_S$, there is only one  $X\in \mathcal{V}$ such that $\mathrm{Tr}_E(X)=x$. This allows us to define the 
 linear \textit{assignment map} $\Lambda_S$ as below. We define  $\Lambda_S(\rho_{S}^{(i)})=\rho_{SE}^{(i)}$, $i=1, \ldots, m$. So, for any $x=\sum_{i=1}^m c_i \rho_{S}^{(i)} \in \mathcal{V}_S$, we have
\begin{equation}
\label{eq:9}
\begin{aligned}
\Lambda_S(x)=\sum_{i=1}^m c_i \Lambda_S(\rho_{S}^{(i)})=\sum_{i=1}^m c_i \rho_{SE}^{(i)}=X.
\end{aligned}
\end{equation} 
 $\Lambda_S$ is a Hermitian map, by construction. It is defined on the whole $\mathcal{V}_S$, and even if $m< {(d_S)}^2$, it can be simply extended  to the  whole $\mathcal{L}(\mathcal{H}_S)$ \cite{16}.
 
Although the assignment map $\Lambda_S$, in Eq. \eqref{eq:9}, is only a mathematical tool which acts as the reverse of  $\mathrm{Tr}_E$, it has an important physical consequence.  It allows us to assign to each $\rho_S \in  \mathrm{Tr}_E( \mathcal{D}_{SE}\cap \mathcal{V})$ a $\rho_{SE} \in  \mathcal{D}_{SE}\cap \mathcal{V}$ such that $\mathrm{Tr}_E(\rho_{SE})=\rho_{S}$.
So, 
  for any $\rho_S \in  \mathrm{Tr}_E( \mathcal{D}_{SE}\cap \mathcal{V})$, and any unitary evolution $U$ of the whole system-environment, the reduced dynamics of the system, using Eqs. \eqref{eq:2} and \eqref{eq:9}, is given by
\begin{equation}
\label{eq:10}
\begin{aligned}
\rho_{S}^{\prime}=\mathrm{Tr}_{E} \circ \mathrm{Ad}_{U}  \circ \Lambda_S(\rho_{S})\equiv 
\Phi_S(\rho_{S}). 
\end{aligned}
\end{equation} 
The unitary evolution $U$ and the partial trace $\mathrm{Tr}_{E}$ are CP maps \cite{1}. We have seen that the assignment map $\Lambda_S$ is, in general, Hermitian. 
Therefore, the  \textit{dynamical map} $\Phi_S$ is, in general, a Hermitian 
map.

It is worth noting  that for (the extension  to the whole $\mathcal{L}(\cH_S)$ of)  each linear trace-preserving Hermitian map, as  $\Phi_S$,  there exists an operator sum representation  such that
\begin{equation}
\label{eq:10-1}
\begin{aligned}
\rho_{S}^{\prime}=\sum_{i}e_{i}\,\tilde{E_{i}}\,\rho_{S}\,\tilde{E_{i}}^{\dagger},\ \ \ \sum_{i}e_{i}\,\tilde{E_{i}}^{\dagger}\tilde{E_{i}}=I_{S},
\end{aligned}
\end{equation}
where $\tilde{E_{i}}$ are linear operators on $\cH_S$ and $e_{i}$ are real coefficients \cite{17, 18, 10}.
Only for the special case that all of the coefficients $e_{i}$ in Eq. \eqref{eq:10-1} are positive, then we can define $E_{i}=\sqrt{e_{i}}\,\tilde{E_{i}}$ and so the reduced dynamics of the system is CP, as   Eq. \eqref{eq:3}.
(Also,  for the  Hermitian map $\Lambda_S$, there exists a similar operator sum representation, as Eq.  \eqref{eq:10-1}, with linear operators $\tilde{E_{i}}: \cH_S\rightarrow\cH_S\otimes\cH_E$.)

 \subsection{Reference state}
 
In Ref. \cite{9}, introducing the \textit{reference states} $\omega_{RSE}$  and $\omega_{RS}=\mathrm{Tr}_{E}(\omega_{RSE})$, we have connected the results of \cite{6} and \cite{10}, reviewed in the two previous subsections. $\omega_{RS}$ is defined as \cite{9}
 \begin{equation}
\label{eq:11}
\begin{aligned}
\omega_{RS}=\sum_{l=1}^{m} \frac{1}{m} \vert l_{R}\rangle\langle l_{R}\vert\otimes \rho_{S}^{(l)},
\end{aligned}
\end{equation}
where $ \rho_{S}^{(l)}\in \mathcal{S}^{\prime}_{S}$ and $\lbrace \vert l_{R}\rangle\rbrace$ is an orthonormal basis for the reference Hilbert space $\mathcal{H}_{R}$.
In addition, the reference state $\omega_{RSE}$ is defined as \cite{9} 
\begin{equation}
\label{eq:12}
\begin{aligned}
\omega_{RSE}= \mathrm{id}_R\otimes\Lambda_S (\omega_{RS})
=\sum_{l=1}^{m} \frac{1}{m} \vert l_{R}\rangle\langle l_{R}\vert\otimes \rho_{SE}^{(l)},
\end{aligned}
\end{equation}
where  $ \rho_{SE}^{(l)}\in \mathcal{S}^{\prime}$ is such that $ \mathrm{Tr}_{E}( \rho_{SE}^{(l)})=\rho_{S}^{(l)}$.

 We can construct subspaces $\mathcal{V}_S$ and 
$\mathcal{V}$ as 
 the\textit{ generalized steered sets}, from $\omega_{RS}$ and $\omega_{RSE}$, respectively. We have \cite{9}
\begin{equation}
\label{eq:13}
\begin{aligned}
\mathcal{V}_{S}=\left\lbrace \mathrm{Tr}_{R}[(A_{R}\otimes I_{S})\omega_{RS}] \right\rbrace ,  
\end{aligned}
\end{equation}
and
\begin{equation}
\label{eq:14}
\begin{aligned}
\mathcal{V}=\left\lbrace \mathrm{Tr}_{R}[(A_{R}\otimes I_{SE})\omega_{RSE}] \right\rbrace ,  
\end{aligned}
\end{equation}
where $ A_{R}$ are arbitrary linear operators in $\mathcal{L}(\mathcal{H}_{R})$.

 When $\omega_{RSE}$, in Eq. \eqref{eq:12}, is a Markov state, as Eq. \eqref{eq:7}, i.e., when there exists a CP assignment map, then, using Eq. \eqref{eq:10}, the reduced dynamics $\Phi_S$ is, consequently,  CP.
 
Comparing  Eqs.  \eqref{eq:6} and  \eqref{eq:14}, shows that, for the steered set $\cS$ from the reference state $\omega_{RSE}$ in Eq. \eqref{eq:12}, we have  $\cS \subseteq\mathcal{D}_{SE}\cap \mathcal{V}$. 
 So, when the reduced dynamics, for all  $\rho_{SE} \in   \mathcal{D}_{SE}\cap \mathcal{V}$, is CP, then, from Theorem \ref{thm 1}, we conclude that  $\omega_{RSE}$ is a  Markov state, as Eq. \eqref{eq:8}.
 
In summary, we have \cite{9}: 
\begin{theorem} 
\label{thm 2}
Consider the subspace $\mathcal{V}\subseteq \mathcal{L}(\cH_S\otimes\cH_E)$, in Eq. \eqref{eq:14}. The reduced dynamics of the system, in Eq. \eqref{eq:10}, is CP, for arbitrary $U$ and any $\rho_{SE} \in   \mathcal{D}_{SE}\cap \mathcal{V}$, if and only if the reference state $\omega_{RSE}$ in Eq. \eqref{eq:12}, is a  Markov state, as Eq. \eqref{eq:8}.
\end{theorem}

\section{Markovianity of the reference state and complete positivity of the reduced dynamics} \label{sec: C}

Theorem \ref{thm 2} is based on Theorem \ref{thm 1}. In Ref. \cite{6}, during the proof of Theorem \ref{thm 1}, it has been assumed that, in general, the unitary time evolution 
$U: \cH_{S}\otimes\cH_{E}\rightarrow \cH_{S}^\prime\otimes\cH_{E}^\prime$ is such that the final Hilbert spaces of the system $\cH_{S}^\prime$  and the environment $\cH_{E}^\prime$ may differ from their initial ones, $\cH_{S}$ and $\cH_{E}$,  respectively.  

In Ref. \cite{11}, we have questioned whether the above assumption can be relaxed or not. In other words, if the reduced dynamics, in  Eq. \eqref{eq:10}, is CP, for arbitrary $U: \cH_{S}\otimes\cH_{E}\rightarrow \cH_{S}\otimes\cH_{E}$ and any $\rho_{SE} \in   \mathcal{D}_{SE}\cap \mathcal{V}$, then 
whether we can conclude that the reference state $\omega_{RSE}$ in Eq. \eqref{eq:12}, is a  Markov state, as Eq. \eqref{eq:8}, or not.

In this section, we consider an example, which is, in fact,  example 4 of Ref. \cite{10}, for which we see that, though 
 the reference state is not a Markov state, the reduced dynamics is CP,  for arbitrary $U: \cH_{S}\otimes\cH_{E}\rightarrow \cH_{S}\otimes\cH_{E}$ and any $\rho_{SE} \in   \mathcal{D}_{SE}\cap \mathcal{V}$. 
Therefore,  the assumption of variability of  Hilbert spaces of the system and the environment, during the time evolution $U: \cH_{S}\otimes\cH_{E}\rightarrow \cH_{S}^\prime\otimes\cH_{E}^\prime$, is necessary, for validity of Theorems \ref{thm 1} and \ref{thm 2}.

Assume that the set $\cS^\prime$ is given by $\cS^\prime=\lbrace\rho, \sigma\rbrace$, where $\rho=\frac{1}{d_Sd_E}I_{SE}$ and $\sigma=\vert 1_S \rangle\langle 1_S\vert \otimes\vert 1_E \rangle\langle 1_E\vert$, $\vert 1_S \rangle\in\cH_S$ and  $\vert 1_E \rangle\in\cH_E$. $\mathcal{V}$ is the subspace spanned by $\cS^\prime$, and $\mathcal{V}_S$ is spanned by $\cS^\prime_S=\lbrace \tilde{\rho}=\frac{1}{d_S}I_{S}, \tilde{\sigma}=\vert 1_S \rangle\langle 1_S\vert\rbrace$. $\cS^\prime_S$ is a linearly independent set, as  $\cS^\prime$. So, there is a one to one correspondence between the members of  $\mathcal{V}$  and  $\mathcal{V}_S$.
 Therefore, from Sec. \ref{sec:B-2}, the reduced dynamics $\Phi_S$, in Eq. \eqref{eq:10}, is given by a Hermitian map, as  Eq. \eqref{eq:10-1}, for arbitrary $U$ and any $\rho_{SE} \in   \mathcal{D}_{SE}\cap \mathcal{V}$.
 
 It can be shown simply that the assignment map $\Lambda_S$, in Eq. \eqref{eq:9}, is non-positive on 
$\mathcal{V}_S$ \cite{10}: for $a\geq 0$ and $-\frac{a}{d_S} \leq b < - \frac{a}{d_Sd_E}$, $x=a \tilde{\rho}+ b\tilde{\sigma} \geq 0$, but $\Lambda_S(x)= a \rho +b \sigma\ngeqslant 0$. 
So, any extension of $\Lambda_S$, to the whole $\mathcal{L}(\mathcal{H}_S)$, is also non-positive, at least, on $\mathcal{V}_S$.
 Therefore, we expect that the reference state $\omega_{RSE}$, in Eq. \eqref{eq:12}, is not a Markov state, as Eq. \eqref{eq:8}. In the following, we show this, explicitly. We have
 \begin{equation}
\label{eq:16}
\begin{aligned}
\omega_{RSE}=  \frac{1}{2} (\vert 1_{R}1_{S}1_{E}\rangle\langle 1_{R}1_{S}1_{E}\vert
 + \frac{1}{d_Sd_E} \vert 2_{R}\rangle\langle 2_{R}\vert\otimes I_{SE}).
\end{aligned}
\end{equation}
Note that $ \langle 1_{R} 1_{E}\vert\omega_{RSE}\vert 1_{R} 1_{E}\rangle= \frac{1}{2} \vert1_{S}\rangle\langle 1_{S}\vert$.  If $\omega_{RSE}$ is a Markov state, then, from  Eq. \eqref{eq:8}, we have
 \begin{equation}
\label{eq:17}
\begin{aligned}
 \langle 1_{R} 1_{E}\vert\omega_{RSE}\vert 1_{R} 1_{E}\rangle=\bigoplus_{k}\lambda_{k}\:\eta_{s^{l}_{k}}\otimes\eta_{s^{r}_{k}}=\frac{1}{2} \vert1_{S}\rangle\langle 1_{S}\vert,
\end{aligned} 
\end{equation}
  where $\eta_{s^{l}_{k}}$ and $\eta_{s^{r}_{k}}$ are positive operators on $\cH_{s^{l}_{k}}$ and $\cH_{s^{r}_{k}}$, respectively. Therefore
   \begin{equation}
\label{eq:18}
\begin{aligned}
    \vert1_{S}\rangle=\vert1_{s^{l}_{k_0}}\rangle\vert1_{s^{r}_{k_0}}\rangle,
  \end{aligned} 
\end{equation}  
     where $\vert1_{s^{l}_{k_0}}\rangle\in \cH_{s^{l}_{k_0}}$ and $\vert1_{s^{r}_{k_0}}\rangle\in \cH_{s^{r}_{k_0}}$, for some $k_0$, and other $\eta_{s^{l}_{k}}$ and $\eta_{s^{r}_{k}}$, for $k\neq k_0$, are zero.
 Now,  Eqs.  \eqref{eq:8} and \eqref{eq:18} result that  $ \langle 1_{S}\vert\omega_{RSE}\vert 1_{S}\rangle$ must be  as $\eta_R\otimes\eta_E$, where $\eta_R$ and $\eta_E$ are positive operators on $\cH_R$ and $\cH_E$, respectively. But, from Eq. \eqref{eq:16}, it can be shown easily that  $ \langle 1_{S}\vert\omega_{RSE}\vert 1_{S}\rangle$  cannot be written in a product form $\eta_R\otimes\eta_E$. Therefore, $\omega_{RSE}$, in Eq.  \eqref{eq:16}, is not a Markov state.
 
 Though $\omega_{RSE}$ is not a Markov state, it can be shown that the reduced dynamics, for arbitrary  $U: \cH_{S}\otimes\cH_{E}\rightarrow \cH_{S}\otimes\cH_{E}$, is CP.
Note that, if we do not extend  $\Lambda_S$ to the whole $\mathcal{L}(\mathcal{H}_S)$, the dynamical map $\Phi_S$, in Eq. \eqref{eq:10}, is a map on $\mathcal{V}_S$ . Now, by CP-ness of $\Phi_S$, we mean that there exists an extension of $\Phi_S$ to the whole $\mathcal{L}(\mathcal{H}_S)$, as $\tilde{\Phi}_S$, 
  such that $\tilde{\Phi}_S$ is a completely positive trace-preserving  map, as Eq. \eqref{eq:3}.
 
A simple way of extending $\Phi_S$ is what is called \textit{zero extension} \cite{10}. 
First, we define the orthonormal  projection  $\mathcal{P}:\mathcal{L}(\mathcal{H}_S)\rightarrow \mathcal{V}_S$ (according to the Hilbert-Schmidt inner
product \cite{1}), as below. For any $A\in \mathcal{L}(\mathcal{H}_S)$, we have
 \begin{equation}
\label{eq:19}
\begin{aligned}
    \mathcal{P}(A)=\sum_{i=1}^2 \mathrm{Tr}(P_i A)P_i,
  \end{aligned} 
\end{equation}
where $P_1=\vert1_{S}\rangle\langle 1_{S}\vert$ and $P_2=\frac{1}{\sqrt{d_S-1}}\sum_{j=2}^{d_S} \vert j_{S}\rangle\langle j_{S}\vert$. $\lbrace \vert j_{S}\rangle\rbrace$ is an orthonormal basis for $\cH_S$, including $\vert1_{S}\rangle$.
$ \mathcal{P}$ is CP, as Eq. \eqref{eq:3}, and, for each $x\in \mathcal{V}_S$, we have $ \mathcal{P}(x)=x$.
Now, the zero extension of $\Phi_S$ to the whole $\mathcal{L}(\mathcal{H}_S)$ is 
\begin{equation}
\label{eq:20}
\begin{aligned}
  \tilde{\Phi}_S=\mathrm{Tr}_{E} \circ \mathrm{Ad}_{U}  \circ \Lambda_S\circ  \mathcal{P}.
  \end{aligned} 
\end{equation}
From Eqs. \eqref{eq:10} and \eqref{eq:19}, it is obvious that, for each $x\in \mathcal{V}_S$, we have $ \tilde{\Phi}_S(x)=\Phi_S(x)$. 

In Ref. \cite{10}, by constructing the Choi matrix (operator) \cite{19}, it has been shown that  $\tilde{\Phi}_S$ is CP, for any $U: \cH_{S}\otimes\cH_{E}\rightarrow \cH_{S}\otimes\cH_{E}$. 
Consider the ket $\vert\xi\rangle=\sum_{i=1}^{d_S} \vert i_R\rangle\vert i_S\rangle \in \cH_R\otimes\cH_S$, which is, up to a normalization factor, the maximally entangled state. The Choi matrix, for the map $ \tilde{\Phi}_S$, is  \cite{10}
\begin{equation}
\label{eq:21}
\begin{aligned}
  \mathrm{id}_R\otimes\tilde{\Phi}_S(\vert\xi\rangle\langle\xi\vert)= \vert 1_R\rangle\langle 1_R\vert
  \otimes  \mathrm{Tr}_E(U\sigma U^{\dagger}) \qquad\qquad\qquad \\
  + (I_R-\vert 1_R\rangle\langle 1_R\vert) \otimes\frac{ d_S\mathrm{Tr}_E(U\rho U^{\dagger})-\mathrm{Tr}_E(U\sigma U^{\dagger})}{d_S-1}.
  \end{aligned} 
\end{equation}
When the final Hilbert spaces of the system and the environment are the same as their initial ones, i.e.,  for all $U: \cH_{S}\otimes\cH_{E}\rightarrow \cH_{S}\otimes\cH_{E}$, then $ d_S\mathrm{Tr}_E(U\rho U^{\dagger})=I_S$. So, the Choi matrix is positive, since it is the summation of two positive operators. 
Therefore, $ \tilde{\Phi}_S$ is CP.

According to Theorem \ref{thm 2}, we expect, from the non-Markovianity of the reference state in Eq. \eqref{eq:16}, that there exists, at least, one unitary evolution $U: \cH_{S}\otimes\cH_{E}\rightarrow \cH_{S}^\prime\otimes\cH_{E}^\prime$ for which the reduced dynamics  is non-CP.
Assume  $U_0$ is such that $\cH_{S}^\prime=\cH_{S}\otimes\cH_{E}$,  and $\cH_{E}^\prime$ is a trivial one dimensional Hilbert space. (In fact, this $U_0$ is what  has been used in Ref. \cite{6}, during the proofs of Theorem \ref{thm 3}, below, and, consequently, Theorem \ref{thm 1}.)
Then, the reduced dynamics of the system , for any $\rho_S \in  \mathrm{Tr}_E( \mathcal{D}_{SE}\cap \mathcal{V})$, is given by 
\begin{equation}
\label{eq:22}
\begin{aligned}
\rho_{S^\prime}^{\prime}=\Phi_S(\rho_{S})=\mathrm{Tr}_{E^{\prime}} \circ \mathrm{Ad}_{U_0}  \circ \Lambda_S(\rho_{S})=\Lambda_S(\rho_{S}),
\end{aligned}
\end{equation} 
which is non-positive, since (any extension of) $\Lambda_S$ is non-positive.

Also, note that we have $ d_S\mathrm{Tr}_{E^\prime}(U_0\rho U_0^{\dagger})=\frac{1}{d_E} I_{SE}$, and 
 $\mathrm{Tr}_{E^\prime}(U_0\sigma U_0^{\dagger})=\sigma=\vert 1_S \rangle\langle 1_S\vert \otimes\vert 1_E \rangle\langle 1_E\vert$. So, the second term, on the right hand side of Eq. \eqref{eq:21}, is non-positive. Therefore, the zero extension $\tilde{\Phi}_S$, for $U_0$, is non-CP, too.
 
 Using this fact that when there exists a positive assignment map $\Lambda_S$, then the reference state $\omega_{RSE}$, in Eq. \eqref{eq:12}, is a Markov state \cite{16}, we can summarize the results of this section as below:
\begin{prop}
\label{pro 1}
Consider the subspace $\mathcal{V}\subseteq \mathcal{L}(\cH_S\otimes\cH_E)$, in Eq. \eqref{eq:14}. The reduced dynamics of the system, in Eq. \eqref{eq:10}, is Hermitian, for arbitrary $U$ and any $\rho_S \in  \mathrm{Tr}_E( \mathcal{D}_{SE}\cap \mathcal{V})$.
 If  the reference state $\omega_{RSE}$ in Eq. \eqref{eq:12}, is not a  Markov state, as Eq. \eqref{eq:8}, then there exists, at least, one $U_0: \cH_{S}\otimes\cH_{E}\rightarrow \cH_{S}^\prime\otimes\cH_{E}^\prime$ for which the reduced dynamics is non-positive. 
 But, the non-Markovianity of  $\omega_{RSE}$ does not guarantee the non-CP-ness of the reduced dynamics, when the unitary evolution $U$ is such that $\cH_{S}^\prime=\cH_{S}$ and  $\cH_{E}^\prime=\cH_{E}$.
\end{prop}

Note that Proposition \ref{pro 1} includes a generalization of 
Theorems \ref{thm 1} and  \ref{thm 2}. Theorems \ref{thm 1} and  \ref{thm 2} state that when  $\omega_{RSE}$ is not a Markov state then there exists, at least, one $U_0$ such that the reduced dynamics is non-CP.
But, Proposition \ref{pro 1} states that the  non-Markovianity of  $\omega_{RSE}$ leads to the non-positivity of the reduced dynamics, for, at least, one $U_0$, as Eq.  \eqref{eq:22}, since the non-Markovianity of  $\omega_{RSE}$ results in the non-positivity of the assignment map $\Lambda_S$ \cite{16}.

\section{Non-Markovianity of the reference state and monotonicity of the relative entropy} \label{sec: D}

In Ref. \cite{13}, it has been shown that the relative entropy,  Eq. \eqref{eq:4},  is monotone under positive trace-preserving maps, as Eq. \eqref{eq:5}.
As we have seen in Sec. \ref{sec:B-2}, the dynamical map $\Phi_S$, in Eq. \eqref{eq:10}, is, in general, a Hermitian trace-preserving map. Therefore, the question   arises as to whether the relative entropy is  monotone under Hermitian maps, too, or not.

In this section,  we show that the result of Ref. \cite{13} cannot be generalized to the Hermitian trace-preserving maps, in general. In other words, there exist
 physically admissible processes for which the relative entropy is not monotone.
 
 First, note that, when the system and the environment undergo the unitary time evolution $U$, jointly,  the reference state $\omega_{RSE}$, in Eq. \eqref{eq:12}, evolves as
 \begin{equation}
\label{eq:23}
\begin{aligned}
\omega_{RSE}^\prime=\mathrm{id}_{R}\otimes\mathrm{Ad}_{U} (\omega_{RSE}).
\end{aligned}
\end{equation} 
 This can be
considered as an actual time evolution, for  a tripartite closed quantum system of  reference-system-environment, during which the reference remains unchanged.

From Eqs. \eqref{eq:12} and \eqref{eq:23}, we have
 \begin{equation}
\label{eq:24}
\begin{aligned}
\omega_{RSE}^\prime=\sum_{l=1}^{m} \frac{1}{m} \vert l_{R}\rangle\langle l_{R}\vert\otimes \rho_{SE}^{\prime (l)},
\end{aligned}
\end{equation} 
where $\rho_{SE}^{\prime(l)}=\mathrm{Ad}_{U}(\rho_{SE}^{(l)})$. Therefore, the evolution of $\omega_{RS}$,  in Eq. \eqref{eq:11}, is given by
 \begin{equation}
\label{eq:25}
\begin{aligned}
\omega_{RS}^\prime=\mathrm{Tr}_{E}(\omega_{RSE}^\prime)=
\sum_{l=1}^{m} \frac{1}{m} \vert l_{R}\rangle\langle l_{R}\vert\otimes \rho_{S}^{\prime(l)}  \\
 =\mathrm{id}_{R}\otimes\Phi_S(\omega_{RS})\equiv\Phi_{RS}(\omega_{RS}), \qquad
\end{aligned}
\end{equation} 
where $\rho_{S}^{\prime(l)}=\mathrm{Tr}_{E}(\rho_{SE}^{\prime(l)})$, and $\Phi_{S}$ is given in Eq. \eqref{eq:10}.  
 Note that $\Phi_{S}$ is a Hermitian map, in general, and so is $\Phi_{RS}$.

In addition, from  Eq. \eqref{eq:25}, we have $\omega_{R}^\prime=\mathrm{id}_{R}(\omega_{R})=\omega_{R}$ and  $\omega_{S}^\prime=\Phi_{S}(\omega_{S})$, where $\omega_{R}=\mathrm{Tr}_{S}(\omega_{RS})$, $\omega_{R}^\prime=\mathrm{Tr}_{S}(\omega_{RS}^\prime)$, $\omega_{S}=\mathrm{Tr}_{R}(\omega_{RS})$, 
 and $\omega_{S}^\prime=\mathrm{Tr}_{R}(\omega_{RS}^\prime)$. So, the  evolution of the state $\sigma_{RS}=\omega_{R}\otimes\omega_{S}$ is also given by $\Phi_{RS}$; i.e., $\sigma_{RS}^\prime=\Phi_{RS}(\sigma_{RS})$.
Equivalently,  we can consider the tripartite state $\sigma_{RSE}=\omega_{R}\otimes\omega_{SE}$, where 
 $\omega_{SE}=\mathrm{Tr}_{R}(\omega_{RSE})$, which evolves as Eq. \eqref{eq:23}: $\sigma_{RSE}^\prime=\mathrm{id}_{R}\otimes\mathrm{Ad}_{U} (\sigma_{RSE})$. Now, it can be shown easily that $\sigma_{RS}^\prime=\mathrm{Tr}_{E}(\sigma_{RSE}^\prime)=\Phi_{RS}(\sigma_{RS})$.

Next, using Eq. \eqref{eq:4}, it can be shown that 
\begin{equation}
\label{eq:26}
\begin{aligned}
S(\omega_{RS}\vert\vert\sigma_{RS})=S(\omega_{RS}\vert\vert\omega_R\otimes\omega_{S}) \qquad\qquad \\
=S(\omega_R)+ S(\omega_S)- S(\omega_{RS})  \\
 =I(R:S)_\omega, \qquad\qquad\qquad
\end{aligned}
\end{equation} 
where  $S(\rho)= - \mathrm{Tr}(\rho \mathrm{log} \rho)$ is the von Neumann entropy, and $I(R:S)_\omega$ is the\textit{ mutual information}, for the bipartite state $\omega_{RS}$ \cite{1}. 
Similarly, we have $S(\omega_{RS}^\prime\vert\vert\sigma_{RS}^\prime) =I(R:S)_{\omega^\prime}$.

We want to verify  whether the monotonicity relation,  Eq. \eqref{eq:5}, is also valid for the Hermitian map $\Phi_{RS}$, within its physical domain, or not. We examine the monotonicity for the two states $\omega_{RS}$ and $\sigma_{RS}$. So, using Eq. \eqref{eq:26}, we want to verify whether
\begin{equation}
\label{eq:27}
\begin{aligned}
I(R:S)_{\omega}\geq I(R:S)_{\omega^\prime}.
\end{aligned}
\end{equation}
The following theorem, proven in Ref. \cite{6}, will be helpful:
\begin{theorem} 
\label{thm 3}
Consider the tripartite state $\omega_{RSE}$, which evolves as Eq. \eqref{eq:23}.
 The inequality \eqref{eq:27}, for the bipartite state  $\omega_{RS}$, holds, for arbitrary $U: \cH_{S}\otimes\cH_{E}\rightarrow \cH_{S}^\prime\otimes\cH_{E}^\prime$, 
 if and only if  $\omega_{RSE}$ is a  Markov state, as Eq. \eqref{eq:8}.
\end{theorem}

Theorem \ref{thm 3} states that when $\omega_{RSE}$ is not a  Markov state, e.g.,  Eq. \eqref{eq:16}, 
 then there exists, at least, one $U$, for which the inequality  \eqref{eq:27} is violated.
 In other words, there exists, at least, one Hermitian map $\Phi_{RS}$, for which we have
\begin{equation}
\label{eq:28}
\begin{aligned}
S(\Phi_{RS}(\omega_{RS})\vert\vert \Phi_{RS}(\sigma_{RS}))= I(R:S)_{\omega^\prime}  \qquad \\
  > I(R:S)_{\omega} 
=S(\omega_{RS}\vert\vert \sigma_{RS}).
\end{aligned}
\end{equation}
Therefore, the relative entropy is not monotone, under Hermitian maps, in general.

Let us illustrate Eq. \eqref{eq:28}, using the example considered in the previous section. Assuming that the system-environment evolution is given by $U_0$, using Eqs. \eqref{eq:11},  \eqref{eq:12}, and  \eqref{eq:22}, we can easily show that
\begin{equation}
\label{eq:29}
\begin{aligned}
\Phi_{RS}(\omega_{RS})=\omega_{RSE}, \qquad\qquad\qquad \\
 \Phi_{RS}(\sigma_{RS})=\sigma_{RSE}= \omega_{R}\otimes \omega_{SE}. \,\,
\end{aligned}
\end{equation}
So, as Eq. \eqref{eq:26},  
\begin{equation}
\label{eq:30}
\begin{aligned}
S(\Phi_{RS}(\omega_{RS})\vert\vert \Phi_{RS}(\sigma_{RS}))=S(\omega_{RSE}\vert\vert \omega_{R}\otimes \omega_{SE}) \qquad\qquad \\
=S(\omega_R)+ S(\omega_{SE})- S(\omega_{RSE}). \,
\end{aligned}
\end{equation}
Now, from Eqs. \eqref{eq:26} and \eqref{eq:30}, we have
\begin{equation}
\label{eq:31}
\begin{aligned}
S(\Phi_{RS}(\omega_{RS})\vert\vert \Phi_{RS}(\sigma_{RS}))-S(\omega_{RS}\vert\vert\sigma_{RS})\qquad \\
=S(\omega_{RS})+ S(\omega_{SE})- S(\omega_{RSE})- S(\omega_{S}). 
\end{aligned}
\end{equation}
The right hand side is always non-negative, using the \textit{strong subadditivity relation} \cite{1}.
In fact, only when $\omega_{RSE}$ is a Markov state, as Eq. \eqref{eq:8}, the right hand side is zero; otherwise, it is greater that zero \cite{15}. So, e.g., for $\omega_{RSE}$ in Eq. \eqref{eq:16}, the inequality \eqref{eq:28} is satisfied, when the evolution of the reference-system-environment is given by $\mathrm{id}_{R}\otimes\mathrm{Ad}_{U_0}$.
For this $\omega_{RSE}$, the right hand side of Eq. \eqref{eq:31} is $0.2375$, when $d_S=d_E=2$.

\section{Non- monotonicity of the relative entropy for a Hermitian evolution which does not change initial Hilbert spaces} \label{sec: E}

In the previous section, we have seen that the result of \cite{13}, of monotonicity of  relative entropy under positive maps, cannot be generalized to Hermitian maps, in general. The example, which we gave, illustrating this result, was for the case that the final Hilbert spaces $\cH_S^\prime$ and $\cH_E^\prime$ differ from their initial ones $\cH_S$ and $\cH_E$, respectively. In this section, we give another example, for which  inequality \eqref{eq:28} is satisfied, while $\cH_S^\prime=\cH_S$ and $\cH_E^\prime=\cH_E$, during the evolution.

We consider the example given in Ref. \cite{20}, in which both the system and the environment are qubits.
An arbitrary  state of the system can be written as
\begin{equation}
\label{eq:32}
\begin{aligned}
\rho_{S}=\frac{1}{2}(I_{S}+ \vec{\alpha}.\vec{\sigma}_{S}),
\end{aligned}
\end{equation}
where $\vec{\sigma}_{S}=(\sigma^{(1)}_{S}, \sigma^{(2)}_{S}, \sigma^{(3)}_{S})$, $\sigma^{(i)}_{S}$ are the Pauli operators, and the \textit{Bloch vector} $\vec{\alpha}=(\alpha^{(1)}, \alpha^{(2)}, \alpha^{(3)})$ is a real
 three dimensional
 vector such that $\vert\vec{\alpha}\vert\leq 1$ \cite{1}.

Consider the following (linear trace-preserving) Hermitian assignment map $\Lambda_{S}$:
\begin{equation}
\label{eq:33}
\begin{aligned}
\Lambda_{S}(\sigma_{S}^{(i)})=\frac{1}{2}\sigma_{S}^{(i)}\otimes I_{E} \qquad  (i=1,\, 2,\,  3),\\
\Lambda_{S}(I_{S})=\frac{1}{2}\left( I_{SE}+a\sum_{i=1}^{3} \sigma^{(i)}_{S}\otimes \sigma^{(i)}_{E}\right),
\end{aligned}
\end{equation}
where $a$ is a fixed real constant. So,
\begin{equation}
\label{eq:34}
\begin{aligned}
\tau_{SE}\equiv\Lambda_{S}(\rho_{S})\qquad\qquad\qquad\qquad\qquad\qquad\qquad\qquad\qquad\\
=\frac{1}{4}\left(I_{SE}+\sum_{i=1}^{3}\alpha^{(i)}\sigma_{S}^{(i)}\otimes I_{E}
 +a\sum_{i=1}^{3} \sigma^{(i)}_{S}\otimes \sigma^{(i)}_{E}\right). \,
\end{aligned}
\end{equation}
When $a\geq 0$, $\tau_{SE}$ is positive for $\vert\vec{\alpha}\vert\leq \sqrt{(1+a)(1-3a)}$, and when $a\leq 0$,  $\tau_{SE}$ is positive for $\vert\vec{\alpha}\vert\leq (1+a)$ \cite{20, 10}. Therefore, for $a\neq 0$,  $\Lambda_{S}$ is a  non-positive map. 

The reference state $\omega_{RSE}$, for this example, is constructed in \cite{9}:
\begin{equation}
\label{eq:35}
\begin{aligned}
\omega_{RSE}=\sum_{l=1}^{3} \frac{1}{16} \vert l_{R}\rangle\langle l_{R}\vert \qquad\qquad\qquad\qquad\qquad\qquad\qquad\\
\otimes \left( I_{SE}+ \alpha^{(l)}\sigma_{S}^{(l)}\otimes I_{E}
 +a\sum_{i=1}^{3} \sigma^{(i)}_{S}\otimes \sigma^{(i)}_{E}\right) \\
 + \frac{1}{16} \vert 4_{R}\rangle\langle 4_{R}\vert\otimes (I_{SE}
 +a\sum_{i=1}^{3} \sigma^{(i)}_{S}\otimes \sigma^{(i)}_{E}),\qquad\quad
\end{aligned}
\end{equation}
where $\alpha^{(l)}$ are arbitrary real constants such that, for $a\geq 0$,  $0< \vert \alpha^{(l)}\vert\leq \sqrt{(1+a)(1-3a)}$, and for $a\leq 0$,  $0< \vert\alpha^{(l)}\vert\leq (1+a)$.
From the non-positivity of the assignment map $\Lambda_{S}$, in Eq. \eqref{eq:33}, we expect that the reference state  $\omega_{RSE}$ is non-Markovian. In \cite{9}, it has been shown that  $\omega_{RSE}$, in Eq. \eqref{eq:35}, is not a Markov state, as Eq. \eqref{eq:8}.

\begin{figure}
\begin{center}
\includegraphics[width=8.75cm]{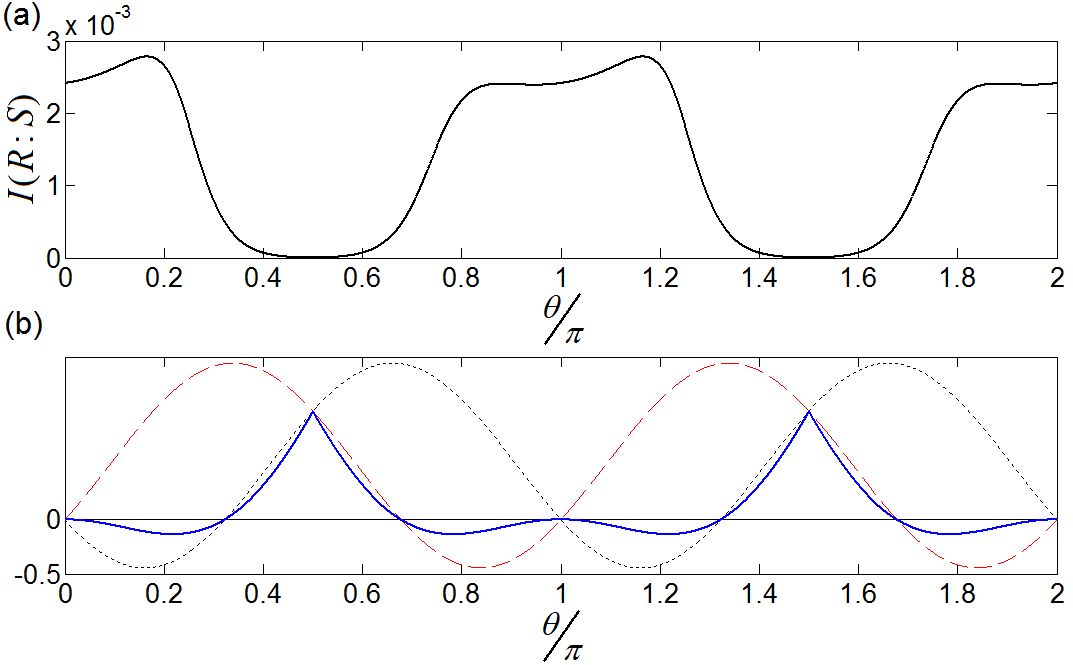}
\end{center}
\caption{(a) Mutual information $I(R:S)_{\omega(\theta)}$, as a function of $\theta$, for  $a=-0.8$,      $\alpha^{(1)}=0.15b$, $\alpha^{(2)}=0.25b$, and $\alpha^{(3)}=-0.6b$,
 where $b=1+a$, in Eq. \eqref{eq:35}.
(b) Three of the eigenvalues of the Choi matrix, which are negative, for some values of  $\theta$,  for $a=-0.8$.}
\label{Fig1}
\end{figure}

According to Theorem \ref{thm 2}, the non-Markovianity of  $\omega_{RSE}$  results in existence of, at least, one $U$, for which the reduced dynamics $\Phi_S$, in Eq. \eqref{eq:10}, is non-CP.
In Ref. \cite{20}, a class of unitary evolutions of the system-environment, as
\begin{equation}
\label{eq:36}
U(\theta)=\left( 
\begin{matrix} 
1&0&0&0\\
0 & \mathrm{cos}\, \theta & \mathrm{sin}\, \theta & 0 \\
0 & - \mathrm{sin}\, \theta & \mathrm{cos}\, \theta & 0 \\
0 & 0 & 0 & 1
\end{matrix}
\right), 
\end{equation}
 has been introduced, where, for some values of $\theta$, the reduced dynamics of the system $\Phi_S(\theta)=\mathrm{Tr}_{E} \circ \mathrm{Ad}_{U(\theta)}  \circ \Lambda_S$ is non-CP \cite{20, 10}.
 The non-CP-ness of $\Phi_S(\theta)$ can be detected by calculating the eigenvalues of the Choi matrix  of it. For this example, the Choi matrix is given explicitly in \cite{10}. When, at least, one of the eigenvalues of the Choi matrix is negative, then $\Phi_S(\theta)$ is non-CP.  For this example, the eigenvalues of the Choi matrix can be calculated analytically. 
In Fig.\ref{Fig1}.b, three of the eigenvalues of the Choi matrix, which are negative, for some values of  $\theta$, are plotted, for $a=-0.8$. (The fourth one is always positive.)

Non-CP-ness of $\Phi_S(\theta)$ results in non-positivity of $\Phi_{RS}(\theta)=\mathrm{id}_{R}\otimes\Phi_S(\theta)$, since $d_R=4>2=d_S$.
From Eq. \eqref{eq:25}, we have $\omega_{RS}(\theta)=\Phi_{RS}(\theta)[\omega_{RS}]$, where $\omega_{RS}=\mathrm{Tr}_{E}(\omega_{RSE})$, and $\omega_{RSE}$ is given in Eq. \eqref{eq:35}.
Fortunately, for this example, the eigenvalues of $\omega_{RS}(\theta)$ and $\omega_{S}(\theta)=\mathrm{Tr}_{R}[\omega_{RS}(\theta)]$ can be calculated analytically. Therefore, from Eq. \eqref{eq:26},
 $I(R:S)_{\omega(\theta)}$, where $\omega(\theta)=\omega_{RS}(\theta)$ can, also, be calculated analytically.
In  Fig.\ref{Fig1}.a, the mutual information $I(R:S)_{\omega(\theta)}$ is plotted as the function of $\theta$. 
Fig. \ref{Fig1}.a shows that  $I(R:S)_{\omega(\theta)}$ exceeds its initial value, for some values of $\theta$. So, for these values of $\theta$, the inequality \eqref{eq:28} is satisfied.
Note that the unitary evolution $U(\theta)$, in Eq. \eqref{eq:36}, does not change $\cH_S$ and $\cH_E$.

Let us summarize the result of the two last sections:
\begin{prop}
\label{pro 2}
The result of \cite{13}, of monotonicity of the relative entropy under positive trace-preserving maps, cannot be generalized to the  Hermitian trace-preserving non-positive maps, within their physical domains, in general.  Inequality \eqref{eq:28} can be satisfied, both when $\cH_{S}$ and  $\cH_{E}$ vary, during the non-positive evolution $\Phi_{RS} =\mathrm{id}_{R}\otimes\Phi_S$, and when they do not vary.
\end{prop}

To achieve the above result, first, we have considered the time evolution of reference-system-environment as Eq. \eqref{eq:23}, which allows us to use Theorem \ref{thm 3}.
 Second, we have considered the two appropriate states  $\omega_{RS}$ and  $\sigma_{RS}$, for which we can write Eq. \eqref{eq:26}, both before and after the evolution  $\Phi_{RS} =\mathrm{id}_{R}\otimes\Phi_S$. Therefore, we could write the monotonicity relation, Eq. \eqref{eq:5}, as the inequality 
 \eqref{eq:27}, which, from Theorem \ref{thm 3},  we know  is violated for a non-Markovian  $\omega_{RSE}$, for, at least, one $U$.

Note that  $\sigma_{S}=\mathrm{Tr}_R(\sigma_{RS})=\omega_{S}$ and, so, $\sigma_{S}^\prime=\Phi_S(\sigma_{S})=\omega_{S}^\prime$. Therefore, for the two equal states $\omega_{S}$ and  $\sigma_{S}$ (and the evolution $\Phi_S$) the monotonicity relation, Eq. \eqref{eq:5}, is, trivially, satisfied.  But, as we have seen, the evolution $\Phi_{RS} =\mathrm{id}_{R}\otimes\Phi_S$ can lead to the violation of the inequality \eqref{eq:5}, for the two states $\omega_{RS}$ and  $\sigma_{RS}$.

\section{Summary }\label{sec: S}

In Ref. \cite{9}, we have introduced the reference states $\omega_{RSE}$,  Eq. \eqref{eq:12}, and $\omega_{RS}$,  Eq. \eqref{eq:11}. There,  we have used them to  connect the results of \cite{6} and \cite{10}, as reviewed in Sec. \ref{sec: B}.
In this paper, we have given two other results, using these reference states.
 
First, in Sec. \ref{sec: C}, giving an explicit example, we have shown that, even when $\omega_{RSE}$ is not a Markov state, as Eq. \eqref{eq:8}, the reduced dynamics of the system can be  CP, for arbitrary system-environment unitary evolution $U$, which does not change $d_S$ and $d_E$.

This shows that the assumption of variability  of  Hilbert spaces of the system and the environment, during the time evolution $U: \cH_{S}\otimes\cH_{E}\rightarrow \cH_{S}^\prime\otimes\cH_{E}^\prime$, is necessary, for validity of Theorems \ref{thm 1} and \ref{thm 2}.

Second, in Sec. \ref{sec: D}, considering the time evolution of the reference states $\omega_{RSE}$ and 
$\omega_{RS}$, and using Theorem \ref{thm 3}, proven in \cite{6}, we have shown that, when $\omega_{RSE}$ is not a Markov state, then there exists, at least, one Hermitian non-positive map $\Phi_{RS} =\mathrm{id}_{R}\otimes\Phi_S$, for which the inequality \eqref{eq:28} is satisfied.
Note that $\omega_{RS}$ and $\sigma_{RS}$, in Eq. \eqref{eq:28}, are in the physical domain of $\Phi_{RS}$.
Therefore, the relative entropy is not monotone, under  Hermitian non-positive maps, even  within their physical domains, in general. 

When $\omega_{RSE}$  is not a Markov state, any possible assignment map $\Lambda_S$ is non-positive \cite{16}. So, choosing $\Phi_S=\Lambda_S$, as Eq. \eqref{eq:22}, results in a non-positive  $\Phi_{RS}$. In Sec. \ref{sec: D}, we have seen that, at least, for this $\Phi_{RS}$, inequality \eqref{eq:28} is satisfied.

In addition to the above example, which includes changes in $d_S$ and $d_E$ after the evolution, in 
 Sec. \ref{sec: E}, we have given another example, for which  inequality \eqref{eq:28} is satisfied, while $\cH_{S}$ and  $\cH_{E}$ remain unchanged, during the evolution.


\end{document}